\documentclass[twocolumn]{aastex63}
\usepackage{natbib}
\usepackage{amsmath}
\submitjournal{ApJ}

\newcommand\sfing {star-forming }

\newcommand\gc{globular cluster }

\newcommand\gcs{globular clusters }

\newcommand\Ms{M_{\odot}}

\newcommand\hhb{hot hydrogen-burning }

\received{2024 Aug 19}
\revised{2024 Oct 9}
\accepted{2024 Oct 9}

\shorttitle{Globular Clusters in the ($m_{1P}$,$F_{1P}$) and ($m_{2P}$,$F_{1P}$) spaces}
\shortauthors{Parmentier}

\begin{document}

\title{Cracking the relation between mass and 1P-star fraction of globular clusters: \\
II. The masses in 1P and 2P stars as a second tool}

\correspondingauthor{Genevi\`eve Parmentier}
\email{gparm@ari.uni-heidelberg.de}

\author[0000-0002-2152-4298 ]{Genevi\`eve Parmentier}
\affiliation{Astronomisches Rechen-Institut, Zentrum f\"ur Astronomie der Universit\"at Heidelberg, M\"onchhofstr. 12-14, D-69120 Heidelberg, Germany}

\begin{abstract}
Galactic globular clusters contain two main groups of stars, the pristine or 1P stars, and the polluted or 2P stars.  
The pristine-star fraction in clusters, $F_{1P}$, is a decreasing function of the cluster present-day mass, $m_{prst}$.  Paper~I  has introduced a model mapping the region of the $(m_{prst},F_{1P})$ space occupied by clusters, with the cluster mass threshold for 2P-star formation a key building-block.  We now expand this model to the pristine-star fraction in dependence of the pristine- and polluted-population masses.  \citet{mil20} found that $F_{1P}$ anticorrelates more tightly with the polluted-population present-day mass, $m_{2P,prst}$, than with the cluster total mass, $m_{prst}$.  In contrast, $F_{1P}$ anticorrelates poorly with the pristine-population current mass,  $m_{1P,prst}$.  We show the loose anticorrelation between $F_{1P}$ and $m_{1P,prst}$ to result from a roughly constant pristine-population mass among clusters as they start their long-term evolution in the Galactic tidal field.  As for the tight anticorrelation between $m_{2P,prst}$ and $F_{1P}$, it stems from the initially shallow relation between $m_{2P}$ and $F_{1P}$ (see Fig.~\ref{fig:mallf1p}).      
Clusters of the Large and Small Magellanic Clouds (LMC and SMC, respectively) appear to behave unexpectedly with respect to each other.  For a given $F_{1P}$, LMC clusters are more massive than SMC clusters despite their enduring a stronger tidal field (see Fig.~\ref{fig:mag}).  This is opposite to how the Galactic outer- and inner-halo clusters behave (see bottom panels of Figs~\ref{fig:mtof1p}-\ref{fig:m2pf1p}).  The explanation may lie in cluster formation conditions.  Finally, we wonder whether the single-population clusters NGC~419 and Rup~106 formed as multiple-population clusters.    

\end{abstract}


\section{Introduction}\label{sec:intro}

Over the past two decades, our knowlegde of Galactic \gcs has advanced tremendously, with ever growing amounts of spectroscopic and photometric data pouring in.  In spite of this data accumulation, or maybe, because of it, a consistent picture of how the oldest clusters of our Galaxy formed has remained elusive.  
    
Photometric and spectroscopic evidence have shown that the vast majority of Galactic \gcs consist of two main stellar populations \citep[e.g.][]{car09,smo11,muccia12,charb16,and09,pio07,mil15}.  The first population (1P) is made up of stars chemically similar to the halo field stars, i.e. O-rich, Na-poor.  We refer to it as the pristine population.  The stars of the second population (2P) are poorer in oxygen but richer in nitrogen, sodium and helium.  Their abundances indicate that they form out of gas polluted with products from the CNO-, NeNa- and, sometimes, MgAl-, cycles \citep[][]{langer93}, hence their name "polluted population".  So far, no consensus has been reached on either the source of the intracluster pollution or its history \citep[e.g.][and references therein]{baslar18}.  Additionnally, the 1P and 2P populations often display complexities of their own.  For instance, the 2P population sometimes consists of distinct subpopulations \citep[e.g. three 2P subpopulations in NGC~2808;][and references therein]{carlos23}, and 1P stars often show internal metallicity spreads \citep{lardo23,legna24}.  No two multiple-population clusters are alike.  

HST photometric studies in the ultra-violet have sharpened the mapping of Galactic \gcs and of their pristine and polluted populations \citep[e.g. ][]{pio15,mil17,mil20,carlos23}.  In particular,  
the census of analyzed stars per cluster has risen by one-to-two orders of magnitude compared to spectroscopic studies.  As a result, reliable estimates of the number fraction of pristine stars, $F_{1P}$, have become available for almost half of the Galactic globular clusters  \citep{mil17,dondo21}.  

The pristine star fraction $F_{1P}$ is a decreasing function of the present-day cluster mass, $m_{prst}$ \citep[see, e.g., ][]{mil20}.  That is, polluted/2P stars dominate the mass budget of the most massive clusters.  An extreme example is $\omega$~Cen, whose fraction in pristine stars \citep[$F_{1P}\simeq 0.09$;][]{dondo21} is the smallest of all analysed Galactic globular clusters.  In contrast, most low-mass \gcs (say those less massive than $3 \cdot 10^4\,\Ms $) are devoid of 2P stars, i.e they are single-population clusters (see Fig.~7 in \citealt{mil20}; see also Fig.~1 in this contribution).  More massive clusters also present more extreme internal abundance variations \citep{lagioia19,zen19}. 

The cluster mass is obviously a key driver of the multiple-population phenomenon, which suggests a stellar mass threshold for 2P-star formation \citep{car10a,bekki11,mil20}.  In the first paper of this series \citep[][hereafter Paper~I]{par24}, we have implemented this threshold concept to decipher the distribution of Galactic \gcs in the $(m_{prst},F_{1P})$ space.  

Suppose \gcs need to reach a fixed stellar mass threshold $m_{th}$ to start forming polluted/2P stars and that, upon reaching this threshold, they form 2P stars exclusively.  The pristine star fraction of newly-formed \gcs then obeys $F_{1P}(m_{ecl}) = m_{th} / m_{ecl}$, where $m_{ecl}$ is the stellar mass of clusters at the end of their formation (equivalently at the end of their gas-embedded phase).  If,  additionally, we assume that the pristine star fraction stays about constant as clusters age,  $F_{1P}(m_{ecl})$ defines the pristine star fraction at any subsequent cluster evolutionnary stage.  To evolve the embedded-cluster mass $m_{ecl}$ up to an age of 12\,Gyr is then all what needs to be done to predict where present-day \gcs are located in the $(m_{prst},F_{1P})$ space. 
The main results and methodology of Paper~I are summarized in Sec.~2.  

The relation of the pristine star fraction $F_{1P}$ with the present-day cluster mass is not the only interesting relation.  Equally relevant are the relations between $F_{1P}$ and the mass of each cluster population.  The pristine star fraction anticorrelates more tightly with the polluted-population present-day mass, $m_{2P,prst}$, than with the present-day cluster mass $m_{prst}$.  On the contrary, $F_{1P}$ anticorrelates only poorly with the pristine-population current mass, $m_{1P,prst}$ \citep[][their Fig.~7]{mil20}.  We shall now expand what has been done in Paper~I to understand these contrasting behaviors, and see how they have been inherited from the cluster initial conditions.  

The outline of the paper is as follows.  Section~\ref{sec:mtof1p} summarizes the results of Paper~I, and update our cluster data set.  Sections \ref{sec:m1pf1p} and \ref{sec:m2pf1p} compare the observed data set in the  $(m_{1P},F_{1P})$ and $(m_{2P},F_{1P})$ spaces with our 12\,Gyr-old model tracks.  Section~\ref{sec:nocor} presents a graphical method that allows us to grasp why a $({\rm mass}, F_{1P})$ data set is strongly or loosely anticorrelated.  It also explores qualitatively how correlation coefficients get altered when the assumption of a fixed stellar mass threshold for 2P-star formation is relaxed.  Section~\ref{sec:mag} shows that, in the $(m_{prst},F_{1P})$ space, clusters of the Small Magellanic Cloud (SMC) and of the Large Magellanic Cloud (LMC) behave unexpectedly with respect to each other. What does it tell us about their formation conditions?  In Sec.~\ref{sec:spmp}, we speculate that the most massive single-population clusters in the Galactic halo (Rup~106) and in the SMC (NGC~419) formed as multiple-population clusters.
We summarize and conclude in Sec.~\ref{sec:conclu}.

\section{Milestones of Paper~I and the $(\lowercase{m_{cluster}},F_{1P})$ space}
\label{sec:mtof1p}
\subsection{Model: summary and key assumptions}
\label{ssec:model}

\begin{figure}
\includegraphics[width=0.49\textwidth, trim = 0 60 0 0]{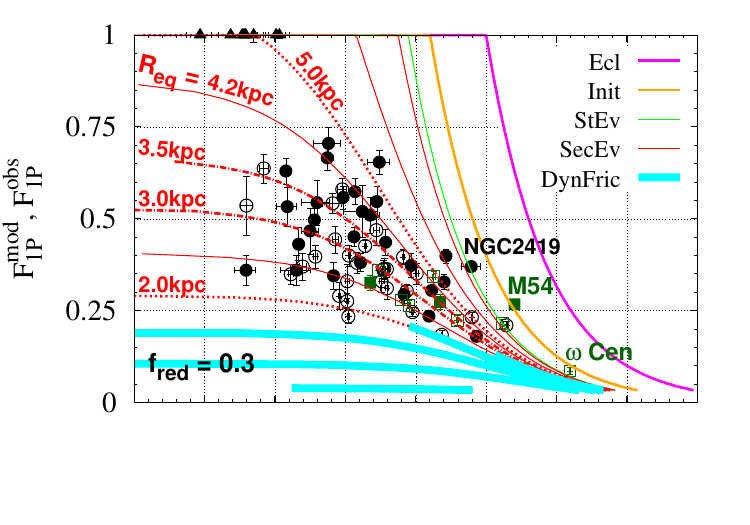}\\
\includegraphics[width=0.49\textwidth, trim = 0  0 0 0]{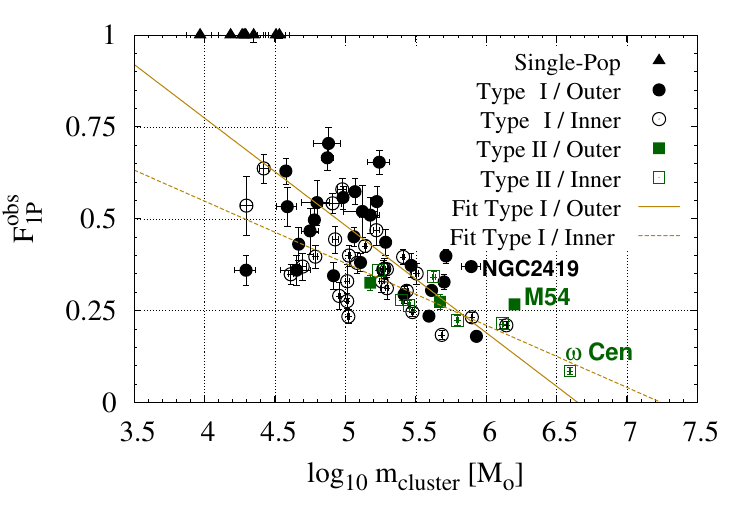}
\caption{Top panel: Relation between cluster mass and pristine-star fraction, with symbols representing present-day Galactic globular clusters, and lines representing model tracks at various cluster evolutionary stages.  Triangles, circles and squares depict single-population clusters, Type~I and Type~II multiple-population clusters, respectively.  Open and plain symbols represent inner ($R_{eq}\leq 3.1$\,kpc) and outer ($R_{eq}>3.1$\,kpc) clusters.  The magenta, orange and green tracks show the pristine-star fraction versus cluster mass at the end of cluster formation, violent relaxation, and stellar evolutionary mass losses, respectively.  NGC~2419, a remote and massive globular cluster, is used to "anchor" the green and orange tracks, assuming 30\,\% stellar evolution mass losses.  Red tracks depict 12\,Gyr-old relations for cluster orbital equivalent radii $R_{eq} = (1-e)D_{apo}=30.0, 8.0, 5.0, 4.2, 3.5, 3.0, 2.5, 2.0, 1.5, 1.0 {\rm ~~and~~} 0.5$\,kpc (the 30\,kpc-track is the one closest to the green track).  Short-dashed lines mark the tracks for $R_{eq}=$2.0 and 5.0\,kpc, and dashed-dotted lines for $R_{eq}=$3.0 and 3.5\,kpc.  Thick cyan lines highlight the region of the diagram where dynamical friction removes massive clusters.  Cluster mass losses affect equally $1P$ and $2P$ stars (i.e. $F_{1P}$ stays constant as clusters evolve).   All model parameters are as in Paper~I.
Bottom panel: Data points overlaid with the least-squares fits for inner (dashed line) and outer (solid line) Type~I clusters }
\label{fig:mtof1p} 
\end{figure}

Paper~I builds on three key assumptions.  Firstly, gas-embedded clusters start forming 2P stars once their stellar mass has reached a fixed threshold $m_{th}$.   Secondly, upon reaching this threshold, the formation of 1P stars is halted  and clusters form 2P stars exclusively.  In other words, the pollution of the intracluster gas in \hhb products, once it starts, is instantaneously completed.  These two assumptions imply that the mass in pristine stars of a newly-formed multiple population cluster is always $m_{th}$.  Let $m_{ecl}$ be the stellar mass of clusters at the end of their  formation (i.e. straight before the winds and ionizing radiations of massive stars remove the residual embedding gas).  Their pristine-star fraction obeys
\begin{equation}
\begin{split}
F_{1P}(m_{ecl})= & \frac{m_{th}}{m_{ecl}} {\rm ~~if~~} m_{ecl}>m_{th}\,, \\
               = & ~~ 1 {\rm ~~~~~~otherwise.}
\label{eq:f1p-ecl}
\end{split}
\end{equation}  
Under Eq.~\ref{eq:f1p-ecl}, a newly-formed cluster of stellar mass $m_{ecl}>m_{th}$ consists of two populations: a pristine-star population of mass $m_{1P,ecl} = F_{1P} m_{ecl} = m_{th}$, and a polluted-star population of mass $m_{2P,ecl} = (1- F_{1P}) m_{ecl} = m_{ecl}-m_{th}$.  Clusters less massive than $m_{th}$ remain single-population clusters (i.e. $F_{1P}=1$).  
Equation~\ref{eq:f1p-ecl} is shown in Fig.~\ref{fig:mtof1p} as the thick magenta line for a mass threshold $m_{th}=10^6\,\Ms$.  We shall discuss this value later in this section.  On the $x$-axis of Fig.~\ref{fig:mtof1p} and in what follows, $m_{cluster}$ is the cluster mass at a specific evolutionary stage (i.e. $m_{cluster}=m_{ecl}$ for the magenta track, while - see below - $m_{cluster}$ is the initial cluster mass for the orange track and the present-day cluster mass for the red tracks).     

To model the region occupied by Galactic \gcs in the $(m_{prst},F_{1P})$ space, Eq.~\ref{eq:f1p-ecl} is evolved up to an age of 12\,Gyr.  Our third hypothesis comes into play here.  We assume that neither of the two populations form more centrally concentrated than the other, thereby implying the constancy of $F_{1P}$ as clusters evolve.  \citet{lei23} have indeed detected dynamically-young \gcs inside which the 1P and 2P populations are similarly distributed (their fig.~15).  
In dynamically-young clusters, 1P and 2P stars have not had the time to mix yet, and their similar present-day distributions must thus be inherited from the formation process.  For such clusters, 1P and 2P stars are lost equally likely, i.e. $F_{1P}$ remains constant as clusters evolve, hence justifying our third hypothesis.  Modeling the evolution of clusters in any $({\rm mass},F_{1P})$ space (where "mass" is either the total mass of clusters, or the mass of their pristine or polluted population) gets therefore reduced to modeling their mass decrease as they age.  Once a cluster bound fraction has been computed, it is equally applied to its total mass, and to the masses of its pristine and polluted populations.  Yet, dynamically-young clusters with a more centrally concentrated population, either 1P or 2P, have also been detected \citep[fig.~15 in][]{lei23}.  In a forthcoming paper, we shall thus relax our third hypothesis by allocating different weights to the fractions of 1P and 2P stars that clusters retain as they evolve. 

We stress that the assumption of 1P and 2P stars being equally likely to be lost from their natal clusters does not contradict the scarcity of 2P field stars in the Galactic halo  \citep[1-3\,\%;][]{car10a,mar11}.  This is because 2P star formation is restricted to the upper part of the embedded-cluster mass spectrum (i.e. clusters with $m_{ecl}>m_{th}$), which, additionally, gives rise to the most resilient clusters (Section~7 of Paper~I). 

Equation~\ref{eq:f1p-ecl} is evolved up to an age of 12\,Gyr in two steps.  Firstly, all clusters are assumed to retain the same stellar mass fraction $F_{bound}^{VR}$ as they violently relax as a result of their massive stars expelling their residual \sfing gas.  In a second step, we use the formalism of \citet{bm03} to model their long-term evaporation in the tidal field of the Galaxy.  

The hypothesis of a fixed fraction $F_{bound}^{VR}$, irrespective of the cluster mass and environment, is discussed in the Sec.~2.2 and Appendix~A of Paper~I.  The cluster mass at the end of violent relaxation is then $m_{init}=F_{bound}^{VR} m_{ecl}$.  We refer to $m_{init}$ as the cluster initial mass as it also defines the cluster mass at the beginning of their long-term evolution in the Galactic tidal field.  The relation between initial mass and pristine-star fraction obeys
\begin{equation}
\begin{split}
F_{1P}(m_{init}) = & F_{1P}(m_{ecl}) = \frac{F_{bound}^{VR} m_{th}}{F_{bound}^{VR} m_{ecl}} \\
                 = & \frac{m_{th,init}}{m_{init}} {\rm ~~if~~} m_{init} > m_{th,init} \\
                 = & ~~ 1 {\rm ~~~~~~~~~~~~~~otherwise.}
\label{eq:f1p-init}
\end{split}
\end{equation}  
Here, $m_{th,init}=F_{bound}^{VR} m_{th}$ is the fixed mass in pristine stars in clusters that have returned to virial equilibrium after gas expulsion.  
$F_{1P}(m_{init})$ is depicted as the orange track in the top panel of Fig.~\ref{fig:mtof1p} (magenta track shifted leftward by an assumed bound fraction $F_{bound}^{VR}=0.40$).  Similarly, the green track accounts for stellar evolutionary mass losses (orange track shifted leftward by an assumed bound fraction $F_{StEv}=0.70$, as appropriate for a canonical IMF).

Prior to modeling the cluster evaporation in the Galactic tidal field, the orange and green tracks must be located in the $({\rm mass},F_{1P})$ space.  
To this end, we adjust the product $F_{StEv} F_{bound}^{VR} m_{th}$ such that the massive and remote, hence little evolved, \gc NGC~2419 sits on the green track.  In other words, we assume that, since the end of its violent relaxation, NGC~2419 has shed in the field the entirety of its stellar evolutionary mass losses, but none of its stars.  For $F_{StEv}=0.70$, this assumption yields $m_{th,init}=F_{bound}^{VR} m_{th} \simeq 4 \cdot 10^5\,\Ms$ (which is directly readable as the intersection between the orange track and the top $x$-axis).  
We coin NGC~2419 our "anchor" cluster, because it allows us to anchor the green and orange tracks in the (mass,$F_{1P}$) space (see Sec.~3 of Paper~I for more details).  

Note that the mass of interest here is the mass in pristine stars at the end of violent relaxation, $m_{th,init}=F_{bound}^{VR} m_{th}$, rather than the mass in pristine stars at the end of cluster formation $m_{th}$.  $m_{th}$ can therefore be decreased as long as $F_{bound}^{VR}$ is increased accordingly.  In Paper~I, we opted for $F_{bound}^{VR}  = 0.4$ and $m_{th}=10^6\,\Ms$.  $m_{th}$ is on the order of the minimum stellar mass needed for clusters to build, via stellar collisions in their inner regions, a super massive star \citep{gie18}.  Super massive stars (stellar mass $\gtrsim 10^4\,\Ms$) are candidate sources of \hhb products \citep{den14}, but have so far remained undetected.  A decrease in $m_{th}$ -- either because $F_{bound}^{VR}$ increases, or because $m_{th,init}$ decreases -- would allow less extreme sources of pollution in \hhb products, e.g. very massive stars \citep{hig23}, which, in contrast to super massive stars, have actually been detected.  In Sec.~\ref{sec:nocor}, we shall see how relaxing the hypothesis of the instantaneous and complete pollution of the intracluster gas (i.e our second hypothesis) decreases $m_{th}$.

The orange and green tracks now depict the starting blocks out of which the 12\,Gyr-old tracks unfold.  To obtain the cluster present-day mass, $m_{prst}$, Paper~I builds on Eqs~10 and 12 of \citet{bm03}.  Their Eq.~12 quantifies the linear decrease of the cluster mass after stellar evolution mass losses, $F_{StEv} m_{init}$, for a given cluster dissolution time-scale.  Equation~10 provides their cluster dissolution time-scale, $t_{diss}^{BM03}$.    $t_{diss}^{BM03}$ scales with $(1-e)D_{apo}$, where $e$ and $D_{apo}$ are the eccentricity and apocentric distance of the cluster orbit.  Paper~I thus defines an equivalent orbital radius $R_{eq}=(1-e)D_{apo}$.  It quantifies the strength of the Galactic tidal field experienced by \gcs (the smaller the apocentric distance and/or the higher the eccentricity, the stronger the external tidal field and the cluster tidal stripping).  The use of this equivalent radius $R_{eq}$ simplifies the comparison between observational data and model tracks as apocentric distances and orbital eccentricities do not need to be considered separately.  

Finally, we multiply the dissolution time-scale of \citet{bm03} by a reduction factor $f_{red} \leq 1$, i.e. $f_{red} t_{diss}^{BM03}$ (Eq.~4 in Paper~I), to account for effects they do not consider and that shorten the life expectancy of clusters, e.g. primordial mass segregation and/or a top-heavy IMF.  We remind the reader that all cluster fractional mass losses are equally applied to their pristine and polluted populations (i.e. $F_{1P}$ stays constant as clusters evolve).    

12\,Gyr-old tracks for $f_{red}=0.3$ and equivalent orbital radii $R_{eq}=30.0, 8.0, 5.0, 4.2, 3.5, 3.0, 2.5, 2.0, 1.5, 1.0, 0.5$\,kpc are shown as the red lines in the top panel of Fig.~\ref{fig:mtof1p} (see also Fig.~5 in Paper~I).  The choice of $f_{red}=0.3$ is discussed in Sec.~3 of Paper~I (see also below).  The red tracks for $R_{eq}=0.5, 1.0, 1.5$ and, to a lesser extent, $2.0$\,kpc, are overlaid with thick cyan lines.  They highlight the region of the diagram where dynamical friction removes massive clusters spiralling toward the Galactic center (Sec.~4.1 of Paper~I).  Outer-halo tracks (i.e.~tracks with large $R_{eq}$) remain in the vicinity of the post-stellar evolution mass loss (green) track.  They correspond to a weak tidal field and to mild cluster evaporation.  In contrast, inner-halo tracks stretch themselves to the left.  They correspond to a stronger tidal field and to greater cluster mass losses.  Star clusters are thus expected to segregate as a function of $R_{eq}$, with outer-halo clusters staying -- on average -- to the right of their inner-halo counterparts (Sec.~5 of Paper~I).  Having summarized the model, let us consider the cluster data set. \\

\subsection{Data set: summary and updates}
\label{ssec:dataset}
     
The pristine-star fractions $F_{1P}^{obs}$ for the clusters of Paper~I are taken from \citet[][their Table~2]{mil17}, completed and updated by \citet[][their Table~3]{dondo21}.  Paper~I also makes use of the sample of \citet[][their Table 4: NGC~1904, NGC~4147, NGC~6712, NGC~7006, and NGC~7492]{jan22}, and considers the following eight clusters -- Rup~106, Ter~7, AM~1, Eridanus, Pal~3, Pal~4, Pa~14 and Pyxis -- as single-population clusters \citep[i.e. $F_{1P}=1$; see][and references therein]{mil20}.  In this contribution, we add two more multiple-population clusters, Arp-2 and Terzan~8, recently studied by \citet[][their Table~3]{lag24}.  \citet{lag24} also add error bars to the pristine-star fractions of the single-population clusters Rup~106 and Ter~7.    

Our sample now includes 74 Galactic globular clusters, divided into three families: 8 single-population clusters, 56 Type~I and 10 Type~II multiple-population clusters.  In contrast to Type~I clusters, Type~II clusters present additional red sequences in their chromosome maps \citep[][see their table~2 for a list]{mil17}, possibly the result of an internal metallicity spread \citep[e.g.][]{mun21}.  Their formation mechanism may thus differ from that of Type~I clusters.  Should they form via cluster mergers \citep[e.g.][]{bekki16} or inside the core of later on tidally-stripped dwarf galaxies \citep[e.g. M54 and $\omega~Cen$;][]{iba94,maj00,hil00}, they do not fit into our scenario outlined above.  We shall thus consider them separately from their Type~I counterparts.  

Figure~\ref{fig:mtof1p} shows the single-population clusters as black triangles (at $F_{1P}=1$), the Type~I multiple-population clusters as black circles, and the Type~II multiple-population clusters as dark-green squares.  Plain and open symbols depict outer-halo and inner-halo clusters, respectively, the separation limit being $R_{eq}=3.1$\,kpc (see below).  Paper~I uses the cluster present-day masses and orbital parameters of \citet[][]{bau19}.  In this contribution, we use their March 2023 updated values, available at \url{https://people.smp.uq.edu.au/HolgerBaumgardt/globular/}.  Knowledge of the cluster orbit apo- and pericentric distances yields their orbital eccentricities $e$ and equivalent orbital radii $R_{eq}$.    

The median equivalent radius of the 56 Type~I clusters is $R_{eq}=(1-e)D_{apo}=3.1$\,kpc.  Based on this value, we divide the Type~I sample into two subsamples of 28 inner-halo clusters ($R_{eq}<3.1$\,kpc; open circles in Fig.~\ref{fig:mtof1p}) and 28 outer-halo clusters ($R_{eq}>3.1$\,kpc; plain circles in Fig.~\ref{fig:mtof1p}).  Following this criterion, single-population clusters (although considered as distinct from the Type~I clusters) are all outer-halo clusters.  This is expected for initially less massive clusters that have survived to this day (Sec.~5 of Paper~I)  \footnote{The median equivalent radius of Paper~I ($R_{eq}=3.3$\,kpc when Type~II clusters are included, and $R_{eq}=3.4$\,kpc when Type~II clusters are excluded) was larger as a result of taking into account single-population clusters.}.       
 
\subsection{Model versus data in the $(m_{cluster}, F_{1P})$ space}
\label{ssec:moddata}

In Paper~I, we show that the 12\,Gyr-old model tracks in the $(m_{prst}, F_{1P})$ space cover the locus of the observational data set provided that the cluster dissolution time-scale of \citet{bm03} is reduced by a factor $f_{red} = 0.3$ (compare figs.~4 and 5 in Paper~I).  This finding may suggest that \gcs start their life either with a top-heavy IMF (Sec.~3.1 in Paper~I), or primordially mass segregated (Sec.~3.2 in Paper~I).  
Section~3.3 in Paper~I suggests yet another possibility: the actual $F_{1P}(m_{init})$ relation is shallower than  assumed by Eq.~\ref{eq:f1p-init}.  A shallower $F_{1P}(m_{init})$ relation would be closer to the data points, thereby requiring longer cluster dissolution time-scales, and a lesser reduction of the \citet{bm03} cluster dissolution time-scale (i.e. $f_{red}$ becomes closer to unity).  A shallower initial track could be the signature of the non-instantaneous pollution of the intracluster gas, and we come back to this in Sec.~\ref{sec:nocor}.

We now obtain the correlation coefficient of the cluster data set.  To compare consistently the correlation coefficients of all three spaces, $(m_{prst}, F_{1P})$, $(m_{1P,prst}, F_{1P})$ and $(m_{2P,prst}, F_{1P})$, we exclude the single-population clusters.  Being devoid of 2P stars, they are absent from the $(m_{2P}, F_{1P})$ space.  We also exclude the Type~II clusters given that their formation mechanism may be more complex than for Type~I clusters (see above).  
When considering the 56 Type~I clusters only, the correlation coefficient is $r_I=-0.56 \pm 0.10$, where the 1-$\sigma$ error has been obtained via the Fischer 's z-transformation \citep{press92}.  That our correlation coefficient is slightly weaker than that found by \citet[][see top-left panel in their Fig.~7: $r=-0.64 \pm 0.08$]{mil20} stems from us excluding the ten Type~II clusters.  Should we consider the Type~I and Type~II clusters together, the correlation coefficient is $r_{I+II}=-0.65 \pm 0.07$, in agreement with \citet{mil20}.  

The bottom panel of Fig.~\ref{fig:mtof1p} shows the cluster data points along with the least-squares fits for the inner (dashed line) and outer (solid line) Type~I clusters: $F_{1P}^{obs,inner}=(-0.17\pm 0.01)\log(m_{prst}^{inner})+(1.22\pm 0.05)$ and $F_{1P}^{obs,outer}=(-0.29\pm 0.01)\log(m_{prst}^{outer})+(1.94\pm 0.07)$.  Our fits slightly differ from those of Paper~I for two reasons: (a) we now exclude the eight single-population clusters, which the Paper~I fits took into account, and (b) we now take into account the $x-$ and $y-$error bars, which were ignored in Paper~I as $F_{1P}$ errors were unkown for all single-population clusters.     
Outer-halo clusters experience smaller mass losses than inner-halo clusters.  This is reflected in their  respective fit locations (see also Sec.~5 of Paper~I). 

In the next two sections, we consider the $(m_{1P}, F_{1P})$ and $(m_{2P}, F_{1P})$ spaces.  It should be noted that since $m_{1P}=F_{1P}m_{cluster}$ and $m_{2P}=(1-F_{1P})m_{cluster}$, the $(m_{1P}, F_{1P})$ and $(m_{2P}, F_{1P})$ spaces do not add new information to Fig.~\ref{fig:mtof1p}; rather, they present it in a different and insightful way.

\section{The $(\lowercase{m}_{1P}, F_{1P})$ space}
\label{sec:m1pf1p}

\begin{figure}
\includegraphics[width=0.49\textwidth, trim = 0 60 0 0]{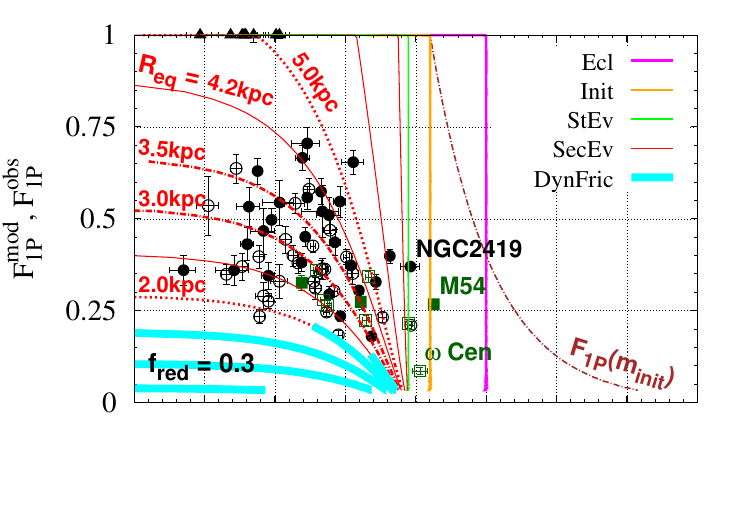}\\
\includegraphics[width=0.49\textwidth, trim = 0  0 0 0]{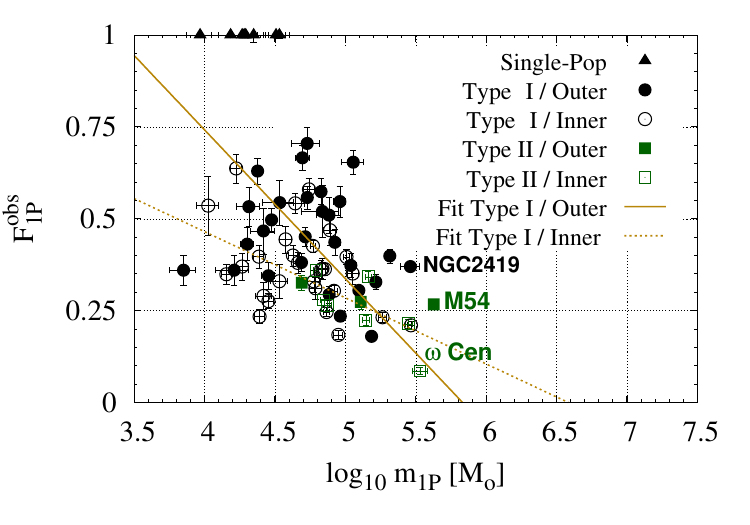}
\caption{Top panel: Relation $F_{1P}(m_{1P})$ between the fraction and mass in pristine stars for models (lines) and Galactic globular clusters (symbols).  Symbol-, line- and color-codings are as in Fig.~\ref{fig:mtof1p}.   The additional dashed-dotted brown line reproduces the pristine-star fraction as a function of the cluster {\it total} initial mass, $F_{1P}(m_{init})$ (i.e. solid orange line in Fig.~\ref{fig:mtof1p}). All cluster mass losses are assumed to affect equally $1P$ and $2P$ stars (i.e. $F_{1P}$ stays constant as clusters evolve).  Note the location of our "anchor" cluster NGC~2419 on the green track.  Its pristine-star mass $F_{StEv} F_{bound}^{VR} m_{th}$ marks the mass threshold for 2P star formation ($m_{th}=10^6\,\Ms$, magenta track) subsequently reduced by the mass losses due to violent relaxation ($F_{bound}^{VR}=0.4$, orange track) and stellar evolution ($F_{StEv}=0.7$, green track).  Bottom panel: Data points overlaid with the least-squares fits for the Type~I inner (dashed lines) and outer (solid lines) clusters  }\label{fig:m1pf1p} 
\end{figure}

Figure~\ref{fig:m1pf1p} presents the pristine-star fraction $F_{1P}$ in dependence of the pristine-population mass $m_{1P} = F_{1P} m_{cluster}$ for the observed sample and our model tracks.  Line-, symbol- and color-codings are as in Fig.~\ref{fig:mtof1p}.  The correlation coefficient for the 56 Type~I clusters is $r_I=-0.26 \pm 0.13$ (for the Type~I and Type~II clusters together, $r_{I+II}=-0.40 \pm 0.11$).  In contrast to the $(m_{prst},F_{1P})$ space, the data here are barely correlated.  To understand why, let us examine how model tracks unfold.  

In the top panel of Fig.~\ref{fig:m1pf1p}, the magenta vertical line is the track at the end of cluster formation.  It corresponds to the mass of the pristine population in newly-formed multiple-population clusters, i.e. $m_{1P,ecl}=m_{th}=10^6\,\Ms$.  
The evolutionary tracks that follow violent relaxation and stellar evolution mass losses are both vertical given the assumed constancy of $F_{bound}^{VR}$ and $F_{StEv}$, among clusters as well as in-between their 1P and 2P populations.  The equation of the (orange) initial track is
\begin{equation}
m_{1P,init} = m_{th,init} = F_{bound}^{VR} m_{th} = {\rm constant}\;.
\label{eq:m1pf1p}
\end{equation}
NGC~2419, our "anchor" cluster, sits on the green track: its present-day mass and pristine-star fraction are $m_{prst,NGC~2419}=7.8 \cdot 10^5\,\Ms$ \citep{bau19}\footnote{\url{https://people.smp.uq.edu.au/HolgerBaumgardt/globular/fits/ngc2419.html}} and $F_{1P,NGC~2419}=0.37$ \citep{zen19}.  Its pristine-population mass is thus $0.37 \cdot 7.8 \cdot 10^5\,\Ms = 2.9 \cdot 10^5\,\Ms = F_{StEv} F_{bound}^{VR} m_{th}$. 
  
The top panel of Fig.~\ref{fig:m1pf1p} also reproduces as the dashed-dotted brown line the pristine-star fraction in dependence of the cluster {\it total} initial mass $m_{init}$ (i.e.~the orange line in Fig.~\ref{fig:mtof1p}).  When $F_{1P}=1$, the two relations $F_{1P}(m_{init})$ and $F_{1P}(m_{1P,init})$ coincide.  As the pristine-population fraction $F_{1P}$ grows thin, however, the polluted-population mass  increases and so does the horizontal gap $m_{init}-m_{th,init}$ between the dashed-dotted brown line and the solid orange line.     

The behavior of the 12\,Gyr-old (red) tracks is akin to that in Fig.~\ref{fig:mtof1p}.  For a given equivalent orbital radius $R_{eq}$, clusters with a high pristine star-fraction experience greater mass losses than their low-$F_{1P}$ counterparts owing to their lower initial mass $m_{init}$.  This  differential behavior yields a negative slope for the tracks.  How steep this slope is depends sensitively on $R_{eq}$.  In the Galactic outskirts (weak tidal field), clusters lose only a minor fraction of their stars.  As a result, model tracks for $R_{eq}=$8 and 30\,kpc are steep and do not deviate much from the green track.  
In contrast, closer to the Galactic center, tracks bend themselves to become almost horizontal, reflecting the severe dissolution of clusters in a strong tidal field.  For instance, when $R_{eq}=2.0$\,kpc, only clusters with $F_{1P}<0.3$ survive up to an age of 12\,Gyr (for $f_{red}=0.3$), unless dynamical friction destroys them.  
The horizontal stretching of the tracks toward zero mass for small $R_{eq}$ is common to all three parameter spaces: $(m_{prst}, F_{1P})$ (Fig.~\ref{fig:mtof1p}), $(m_{1P}, F_{1P})$ (Fig.~\ref{fig:m1pf1p}), and $(m_{2P}, F_{1P})$ (Fig.~\ref{fig:m2pf1p} and Sec.~\ref{sec:m2pf1p} below).  Again, it stems from applying the cluster bound fraction to its 1P and 2P populations equally.   

Figure~\ref{fig:m1pf1p} presents a key difference with Fig.~\ref{fig:mtof1p}, however: the initial-track behaviors and, as a result, the outer-halo track behaviors.  The outer-halo (red) tracks in Fig.~\ref{fig:m1pf1p} are much steeper than in Fig.~\ref{fig:mtof1p}.  This combination, from nearly horizontal tracks for the inner halo to nearly vertical tracks for the outer halo, readily explains the cloudy aspect of the data-point distribution and the weakness of their correlation in the ($m_{1P}$,$F_{1P}$) space.  We will come back to this in Sec.~\ref{sec:nocor}.

\begin{figure}
\includegraphics[width=0.49\textwidth, trim = 0 60 0 0]{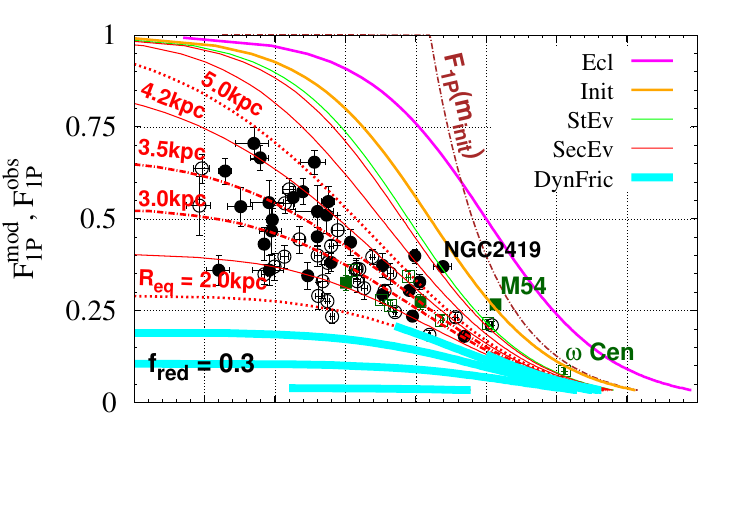} \\
\includegraphics[width=0.49\textwidth, trim = 0  0 0 0]{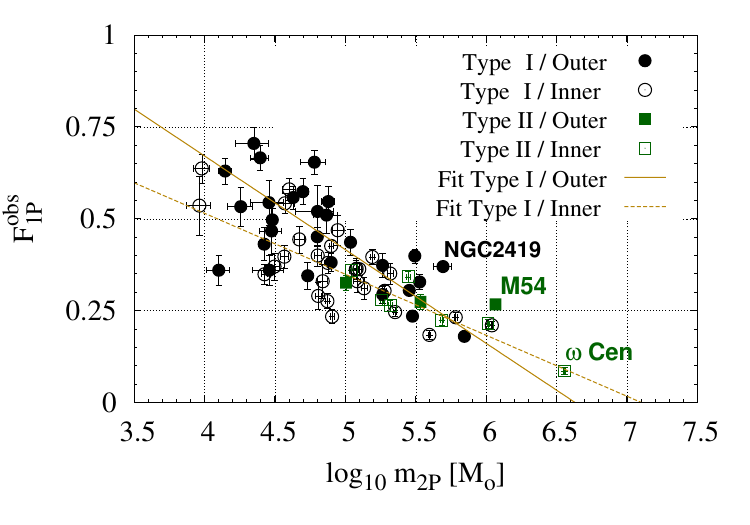}
\caption{Same as Fig.~\ref{fig:m1pf1p} but for $F_{1P}(m_{2P})$. 
}
\label{fig:m2pf1p} 
\end{figure} 

For the sake of completeness, the bottom panel of Fig.~\ref{fig:m1pf1p} highlights how the Type~I inner-halo clusters ($R_{eq} < 3.1$\,kpc) segregate from their outer-halo siblings ($R_{eq} > 3.1$\,kpc).  Their respective  least-squares linear fits are shown as the dotted and solid lines, respectively: $F_{1P}^{obs,inner}=(-0.18\pm 0.01)\log(m_{1P,prst}^{inner})+(1.19\pm 0.07)$ and $F_{1P}^{obs,outer}=(-0.41\pm 0.02)\log(m_{1P,prst}^{outer})+(2.36\pm 0.11)$.  As expected from the  model tracks, the inner clusters fit is shallower and has a smaller intercept than the outer clusters fit.

\section{The $(\lowercase{m}_{2P}, F_{1P})$ space}
\label{sec:m2pf1p}

Of the three relations $F_{1P}^{obs}(m_{prst})$, $F_{1P}^{obs}(m_{1P,prst})$ and $F_{1P}^{obs}(m_{2P,prst})$, \citet{mil20} find $F_{1P}^{obs}(m_{2P,prst})$ to be the one most tightly correlated (correlation coefficient = $-0.77 \pm 0.06$).  Considering only the Type~I clusters of our sample, we indeed find $r_{I}=-0.70 \pm 0.07$ (for the Type~I and Type~II clusters together, $r_{I+II}=-0.76 \pm 0.05$).  We can intuitively understand the tightness of this correlation by extending what we have learned in Secs~\ref{sec:mtof1p} and \ref{sec:m1pf1p}.  The initial track $F_{1P}(m_{2P,init})$ (orange track in Fig.~\ref{fig:m2pf1p}) is here significantly shallower than in the $(m_{cluster},F_{1P})$ and $(m_{1P},F_{1P})$ spaces.  It brings the outer-halo tracks closer to the inner-halo ones, thereby "closing" the region in which cluster data points are distributed.  

Figure~\ref{fig:m2pf1p} is the counterpart of Fig.~\ref{fig:m1pf1p} for the ($m_{2P}$,$F_{1P}$) space.     
When $F_{1P} \gtrsim 0$, the initial mass of the polluted population amounts to almost the cluster total initial mass ($m_{2P,init} \lesssim m_{init}$), and the dashed-dotted brown line and solid orange line almost coincide.  In contrast, when $F_{1P} \lesssim 1$, the scarce 2P population pulls all tracks leftward, the extreme case being $log_{10}(m_{2P}) \rightarrow -\infty$ when $F_{1P}=1$.   The initial/orange track in the ($m_{2P}$,$F_{1P}$) space is thus shallower than in its ($m_{cluster}$,$F_{1P}$) counterpart, which explains the tighter data set correlation. 
Given our definition of the polluted-population mass ($m_{2P,ecl} = m_{ecl} - m_{th}$), the magenta and orange tracks in Fig.~\ref{fig:m2pf1p} obey
\begin{equation}
F_{1P} = \left( 1 + \frac{m_{2P,ecl}}{m_{th}} \right)^{-1} = \left( 1 + \frac{m_{2P,init}}{m_{th,init}} \right)^{-1}\;.
\label{eq:m2pf1p}
\end{equation}
 
The main characteristics of Fig.~\ref{fig:m2pf1p} are otherwise those already highlighted in Figs~\ref{fig:mtof1p} and \ref{fig:m1pf1p}: outer-halo 12\,Gyr-old tracks staying close to the green track, while inner-halo 12\,Gyr-old tracks stretch out horizontally; dearth of clusters both in the region impacted by dynamical friction, and to the left of the green track \citep[exceptions being $\omega$~Cen and M54, consistent with their forming inside larger systems such as dwarf galaxies;][]{maj00,hil00,iba94}.  
Note that single-population clusters are excluded from the ($m_{2P}$,$F_{1P}$) space. 

The bottom panel of Fig.~\ref{fig:m2pf1p} shows the least-squares linear fits for the Type~I inner and outer clusters (dotted and solid lines, respectively):  $F_{1P}^{obs,inner}=(-0.17\pm 0.01)\log(m_{2P,prst}^{inner})+(1.18\pm 0.05)$ and $F_{1P}^{obs,outer}=(-0.26\pm 0.01)\log(m_{2P,prst}^{outer})+(1.69\pm 0.05)$. 
While the inner clusters are slightly shifted to the left with respect to their outer counterparts, as expected, the effect is not as significant as in the ($m_{1P}$,$F_{1P}$) or ($m_{prst}$,$F_{1P}$) space.  This reflects the tighter correlation of the data points in the ($m_{2P}$,$F_{1P}$) space.  It is worth stressing that a least-squares fit in the ($m_{2P}$,$F_{1P}$) space cannot describe the data over the full range of $F_{1P}$ values as a straigthline yields $m_{2P}\ne 0$ when $F_{1P}=1$.  A consistent model must present $log_{10}(m_{2P})$ tending asymptotically toward $-\infty$ as $F_{1P}$ approaches unity, as the orange track in Fig.~\ref{fig:m2pf1p} does.  

\section{Peering into the weak $(\lowercase{m}_{1P,\lowercase{prst}}, F_{1P})$ correlation}
\label{sec:nocor}
\begin{figure} 
\begin{center} 
\includegraphics[width=0.49\textwidth, trim={0.cm 0.7cm 0cm 0cm}]{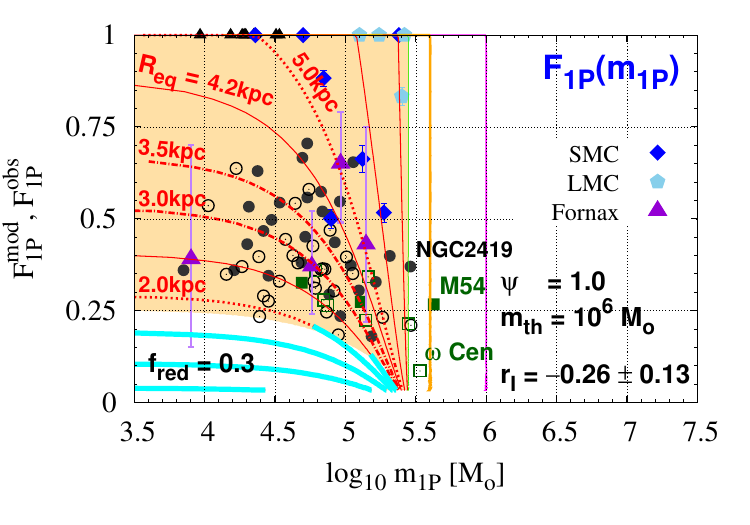} \\
\includegraphics[width=0.49\textwidth, trim={0.cm 0.7cm 0cm 0cm}]{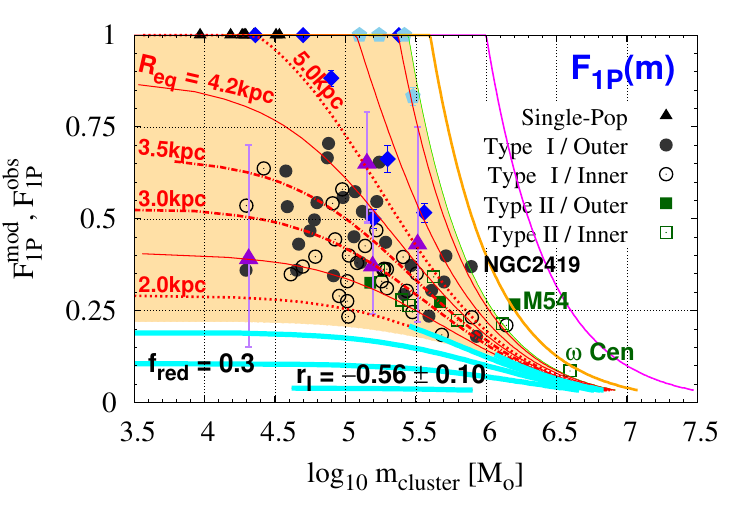} \\
\includegraphics[width=0.49\textwidth, trim={0.cm 0.3cm 0cm 0cm}]{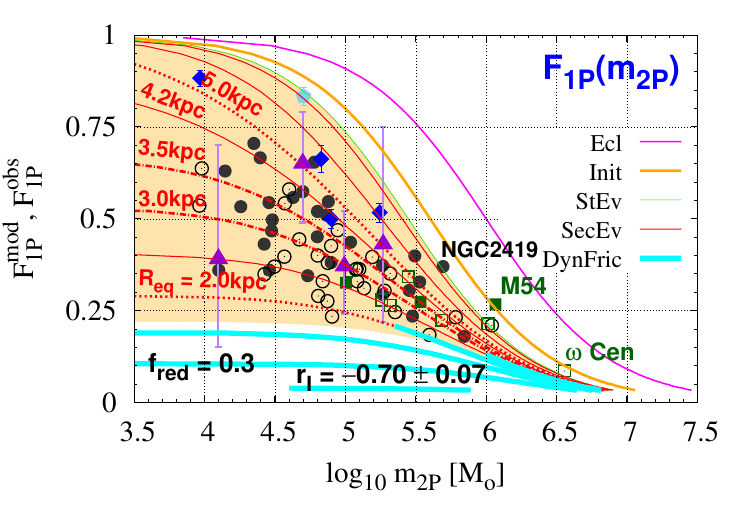} 
\caption{Comparison between model tracks and observed cluster data points.  The permitted region (from the green to the cyan tracks) is highlighted in yellow.  Clusters mapped for multiple-stellar populations in the Magellanic Clouds \citep{mil20,dondo21} and the Fornax dwarf galaxy \citep{lar14} have been added.  Color-, symbol- and line-codings are otherwise as in previous figures.  For the sake of clarity, error bars have been removed for Galactic clusters.  Top panel: ($m_{1P}$,$F_{1P}$) space (akin to top panel of Fig.~\ref{fig:m1pf1p}).  Middle panel: ($m_{cluster}$,$F_{1P}$) space (akin to top panel of Fig.~\ref{fig:mtof1p}).  Bottom panel: ($m_{2P}$,$F_{1P}$) space (akin to top panel of Fig.~\ref{fig:m2pf1p}).  The opening of the permitted region reduces itself from the top through the bottom panel.  It explains why the correlation coefficient for Galactic Type~I globular clusters, $r_{I}$ (quoted at the bottom of each panel), is the poorest in ($m_{1P}$,$F_{1P}$) space and the tightest in the ($m_{2P}$,$F_{1P}$) space.  }
 \label{fig:mallf1p} 
 \end{center} 
\end{figure}

Fig.~\ref{fig:mallf1p} gathers the top panels of Figs~\ref{fig:m1pf1p}, \ref{fig:mtof1p} and \ref{fig:m2pf1p} (in that order, from top to bottom), with a yellow background highlighting the permitted region of each three parameter spaces.  Each permitted region stretches from the green track, namely, the limit imposed by the initial conditions corrected for stellar evolution mass losses, to the cyan tracks, where massive clusters too close to the Galactic center get removed by dynamical friction.  
Figure~\ref{fig:mallf1p} also completes our sample of Galactic \gcs with the sample of Magellanic Clouds clusters of \citet[][updated by \citealt{dondo21}]{mil20} and with the sample of Fornax clusters of \citet{lar14} (see Sec.~6 in Paper~I for more details).  No cluster mass errors are available for those.  For the sake of clarity, Galactic cluster data points are depicted without their error bars.\\

As already mentioned, all three present-day distributions widen toward higher pristine-star fractions, equivalently toward lower embedded-cluster masses.  This stems from the greater mass losses experienced by less massive clusters as they secularly evolve, combined to their greater sensitivity to the external tidal field.  For each parameter space, the opening of the permitted/yellow-highlighted region allows one to gauge the corresponding degree of correlation.
That \gc data are poorly correlated in the ($m_{1P}$, $F_{1P}$) space (Sec.~\ref{sec:m1pf1p}) shows up as the permitted region opening itself up under an angle of almost $90^\circ$ (top panel of Fig.~\ref{fig:mallf1p}).  In contrast, the correlation is at its tightest for $F_{1P}^{obs}(m_{2P})$ (Sec.~\ref{sec:m2pf1p}), and the yellow-highlighted permitted region is thus the narrowest in the bottom panel.  These different configurations, as well as the average model-track slopes, are directly inherited from the initial conditions (orange and green tracks). \\

\begin{figure} 
\begin{center} 
\includegraphics[width=0.49\textwidth, trim={0.cm 0.7cm 0cm 0cm}]{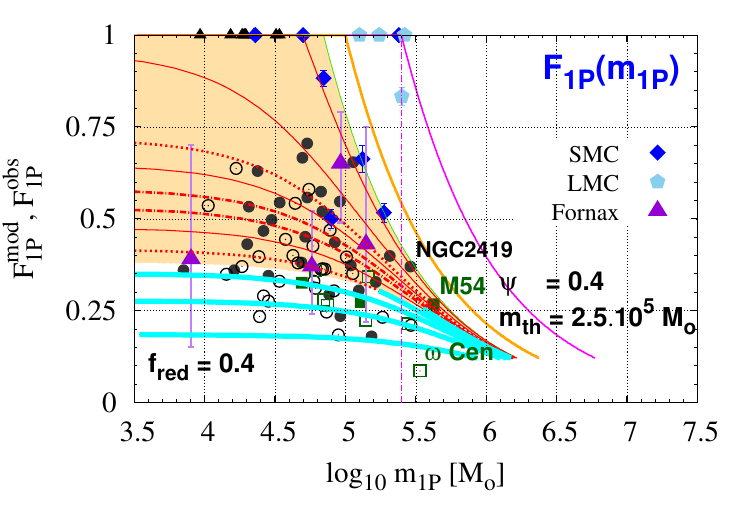} \\
\includegraphics[width=0.49\textwidth, trim={0.cm 0.7cm 0cm 0cm}]{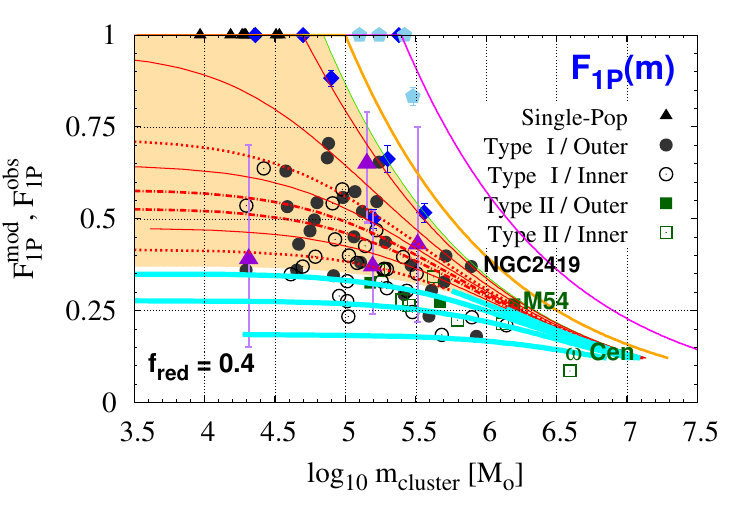} \\
\includegraphics[width=0.49\textwidth, trim={0.cm 0.3cm 0cm 0cm}]{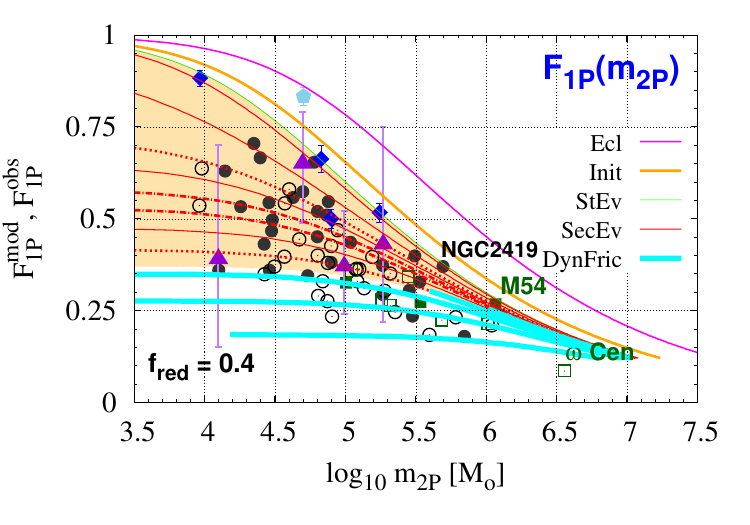} 
\caption{Same as Fig.~\ref{fig:mallf1p} but for a shallower $F_{1P}(m_{ecl})$ relation: $\psi = 0.4$ and $m_{th} = 2.5\cdot 10^5\,\Ms$ in Eq.~\ref{eq:f1p-ecl_sh}}
 \label{fig:mallf1p_sh} 
 \end{center} 
\end{figure}

The most interesting result of Fig.~\ref{fig:mallf1p} is certainly the near-absence of a correlation in the ($m_{1P}$,$F_{1P}$) space (top panel).  As seen in Sec.~\ref{sec:m1pf1p}, this can be explained by the verticality of the initial track.   Should we adopt a significantly shallower initial relation $F_{1P}(m_{1P,init})$, the permitted region would shrink from its top-right-side (see the top panel of Fig.~\ref{fig:mallf1p_sh}), thereby forcing an anticorrelation between the present-day mass $m_{1P,prst}$ and fraction $F_{1P}$.  We shall discuss later in this section the physical implications of a shallower initial relation $F_{1P}(m_{1P,init})$.  Suffice it to say for now that it points either to a cluster-mass dependent threshold for 2P-star formation, or to the non-instantaneous pollution of the forming cluster (i.e.~part of the pristine stars form concurrently with the polluted ones).  We shall also see that these alternative scenarios should not be dismissed, even though, at first sight, the shape of the permitted region in the top panel of Fig.~\ref{fig:mallf1p_sh} contradicts the weak correlation shown by the Type~I \gcs in the $(m_{1P},F_{1P})$ space ($r_I=-0.26_{-0.12}^{+0.13}$, Sec.~\ref{sec:m1pf1p}).  Let us therefore investigate what the corresponding changes in the $(m_{cluster},F_{1P})$ and $(m_{2P},F_{1P})$ spaces are (middle and bottom panels of Fig.~\ref{fig:mallf1p_sh}, respectively).  \\

All three panels of Fig.~\ref{fig:mallf1p_sh} build on the $F_{1P}(m_{ecl})$ relation
\begin{equation}
\begin{split}
F_{1P}(m_{ecl})= & \left( \frac{m_{th}}{m_{ecl}}\right)^\psi {\rm ~~if~~} m_{ecl}>m_{th} \\
               = & ~~ 1 {\rm ~~~~~~otherwise,}
\label{eq:f1p-ecl_sh}
\end{split}
\end{equation}  
with $\psi=0.4$ and $m_{th}=0.25 \cdot 10^6\,\Ms$.  Shown as the magenta track of the middle panel, it is shallower than Eq.~\ref{eq:f1p-ecl} ($\psi=0.4$ instead of $\psi=1.0$), has a lower cluster mass threshold for 2P star formation ($m_{th}=0.25 \cdot 10^6\,\Ms$ instead of $m_{th}= 10^6\,\Ms$), while still retaining NGC~2419 as the "anchor" cluster.  This choice of $\psi$ is for now purely illustrative.  In a forthcoming paper, we shall map with a sounder approach the $\psi$-variations authorized by the observational data.  

We can now understand why the magenta track in the top panel of Fig.~\ref{fig:mallf1p_sh} is shallower than its counterpart of Fig.~\ref{fig:mallf1p} as well.  Combining Eq.~\ref{eq:f1p-ecl_sh} with the definition of the pristine-star fraction in the embedded phase $F_{1P} = m_{1P,ecl} / m_{ecl}$ yields 
\begin{equation}
F_{1P}(m_{1P,ecl}) = \left( \frac{m_{th}}{m_{1P,ecl}} \right)^\frac{\psi}{1-\psi}\;.
\label{eq:m1pf1p_sh}
\end{equation}
Its logarithmic slope when $\psi=0.4$ is $-\psi/(1-\psi)=-2/3$.  When $\psi = 1$, one recovers a slope $\rightarrow -\infty$, as in the top panel of Fig.~\ref{fig:mallf1p}.  As for the pristine-population mass in newly-formed globular clusters,  it obeys $m_{1P,ecl}=F_{1P} \cdot m_{ecl} = m_{th}^{\psi} m_{ecl}^{1-\psi}$.  For $\psi=1$, we recover the fixed pristine-population mass $m_{1P,ecl}=m_{th}$ of the previous sections.  For $0<\psi<1$, $m_{1P,ecl}$ is a growing function of the embedded-cluster mass $m_{ecl}$.

There is no analytical expression for $F_{1P}(m_{2P,ecl})$ (magenta track in bottom panel).  However, the mass of the polluted population in the embedded phase can be expressed as a function of $F_{1P}$:
\begin{equation}
m_{2P,ecl} = m_{th} \frac{1-F_{1P}}{F_{1P}^{1/\psi}}\;.
\end{equation}
If $F_{1P}=1$, $m_{2P,ecl} =0$ (single-population cluster).  If $F_{1P}=0$, $m_{2P,ecl}\rightarrow \infty$, i.e. no pure polluted-population cluster should be able to form since 2P stars cannot form until 1P stars have polluted the intracluster medium.   \\

All yellow-highlighted regions of Fig.~\ref{fig:mallf1p_sh} ($\psi=0.4$) are narrower than their counterparts of Fig.~\ref{fig:mallf1p} ($\psi=1.0$).  We therefore conclude that a shallower $F_{1P}(m_{ecl})$ relation (Eq.~\ref{eq:f1p-ecl_sh} with $\psi<1$) drives tighter correlations in all three spaces ($m_{1P,prst}$,$F_{1P}$), ($m_{prst}$,$F_{1P}$) and ($m_{2P,prst}$,$F_{1P}$). \\

Based on the paucity of the correlation shown by the Type~I \gcs in the $(m_{1P},F_{1P})$ space ($r_I=-0.26_{-0.12}^{+0.13}$, Sec.~\ref{sec:m1pf1p}), one might be tempted to conclude that the data demand a model with $\psi \simeq 1.0$, rather than with $\psi \simeq 0.4$.  It is a step we will not take, however.  The reason is that the modeling in Figs~\ref{fig:mallf1p} and \ref{fig:mallf1p_sh} assumes a one-to-one relation between the initial mass of clusters and their pristine-star fraction (and, therefore, also between the initial mass of their 1P and 2P populations and $F_{1P}$).  That is, no scatter has been taken into account.  Yet, even if the relation between cluster total mass and pristine-star fraction is one-to-one at the end of the embedded stage (magenta tracks), scatter would ensue following violent relaxation.  This is because the response of clusters to gas expulsion depends on a deep bench of parameters (star formation efficiency, gas-expulsion time-scale, virial state prior to gas expulsion, external tidal field in combination to cluster density, respective distributions of the cluster stars and of the residual embedding gas, cluster geometry; see Sec.~2.2 in Paper~I and references therein).  Last but not least, not all clusters retain a constant $F_{1P}$ as they violently relax.  If the polluted population  forms centrally concentrated, pristine stars are the first ones to be stripped unbound, thereby decreasing the pristine-star fraction $F_{1P}$ \citep{ves10}.  Conversely, if the cluster outskirts are preferentially populated by polluted stars, their tidal striping raises the pristine-star fraction $F_{1P}$.  In summary, we do expect the actual $F_{1P}(m_{init})$ relation to show some scatter, which will blur the neat edges of all permitted regions in Figs~\ref{fig:mallf1p} and \ref{fig:mallf1p_sh}.

As a consequence, relations with $0<\psi<1$, like the shallower magenta tracks of Fig.~\ref{fig:mallf1p_sh}, should not be hastily dismissed.  If the initial relations that they yield (orange tracks) come with a scatter, the anticorrelations suggested by the yellow-highlighted regions of Fig.~\ref{fig:mallf1p_sh} will weaken, and the current contradiction with the poorly correlated data in the $(m_{1P},F_{1P})$ space (top panel) may vanish as well.   
Therefore, at this stage, we cannot reject a model that obeys -- {\it on the average} -- Eqs~\ref{eq:f1p-ecl_sh}-\ref{eq:m1pf1p_sh} with $0 < \psi < 1$\footnote{One may also conclude from Fig.~\ref{fig:mallf1p_sh} that the model with $\psi=0.4$ and $m_{th}=0.25 \cdot 10^6\,\Ms$ performs more poorly than with $\psi=1.0$ and $m_{th}=10^6\,\Ms$ (e.g. presence of clusters in the region affected by dynamical friction, model tracks overestimating the cluster pristine-star fraction in the high-mass regime).  An improved modeling of the initial relation when $\psi<1$ is deferred to a forthcoming paper.}.  Let us therefore briefly comment on the physical differences between the models underpinning Figs~\ref{fig:mallf1p} and \ref{fig:mallf1p_sh}.  

As already seen, the model with $\psi = 1$ (Fig.~\ref{fig:mallf1p}) implies a fixed cluster mass threshold for 2P star formation ($m_{th}$), followed by the exclusive formation of 2P stars (i.e. $m_{1P,ecl}=m_{th}$ is a constant and the magenta track in the top panel of Fig.~\ref{fig:mallf1p} is vertical).

When $\psi = 0.4$ (Fig.~\ref{fig:mallf1p_sh}), two extreme scenarios are possible.  A first scenario is that the formation of pristine stars is halted once clusters self-pollute (as previously, the self-pollution of clusters in \hhb products is complete as soon as it starts).  The threshold mass is then $m_{1P,ecl}$, which increases with the embedded cluster mass $m_{ecl}$ ($m_{th}$ becomes the threshold mass only at $F_{1P}\lesssim1$).  In a second scenario, the threshold mass retains a fixed value $m_{th}$ (dashed-dotted vertical line in top panel of Fig.~\ref{fig:mallf1p_sh}), and pristine stars keep forming along with polluted stars (for instance in the cluster outskirts if the \hhb products are given off, or collected, at the cluster center, as is often assumed; e.g. \citealt{der08,gie18}).  In the top panel of Fig.~\ref{fig:mallf1p_sh}, this additional mass in 1P stars shows up as the horizontal shift between the dashed-dotted vertical line, which marks the fixed threshold mass $m_{th}$, and the solid magenta line, which marks the total mass in pristine stars once cluster formation is over.

\section{The weird case of the Magellanic Clouds clusters}
\label{sec:mag}
The Magellanic Clouds clusters shown in Fig.~\ref{fig:mallf1p} are younger than Galactic \gcs \citep[see Table~2 in][]{mil20}.  Additionnally, they evolve in dwarf irregular galaxies, hence in a weak tidal field.  Their resulting limited mass losses mean that they occupy the region of the $(m_{prst}, F_{1P})$ space corresponding to the outer Galactic halo (Sec.~6 of Paper~I and Fig.~\ref{fig:mallf1p}). 

{\it In-between them, however}, the Magellanic Clouds clusters behave unexpectedly: for a given $F_{1P}$, SMC clusters stand to the left of their LMC counterparts\footnote{One exception is NGC~419, an SMC single-population cluster that, in terms of mass, is similar to the LMC single-population clusters.  See also Sec.~\ref{sec:spmp}}.  Yet, should the LMC and SMC clusters share the same initial sequence $F_{1P}(m_{init})$, we would have expected the opposite as a result of the SMC tidal field being weaker than that of the LMC.  That is, we would have expected the SMC clusters to behave with respect to the LMC clusters as the Galactic outer-halo clusters behave with respect to their inner-halo counterparts.  
The SMC is actually about ten times less massive than the LMC \citep{bekki08}, and the SMC should thus provide the more gentle environment.  Estimates of the cluster dissolution time-scale for both dwarf galaxies corroborate this.  For the SMC, \citet{lam05a} estimate the (secular) dissolution time-scale of clusters of initial mass $m_{init} = 10^4\,\Ms$ to be $t_{diss,4} \simeq 8$\,Gyr (where the subscript '4' reminds that this is the dissolution time-scale of a cluster with $m_{init} = 10^4\,\Ms$).  As for the LMC, \citet{par08} model its cluster age and mass distributions to infer a lower limit on the cluster dissolution time-scale, $1\,{\rm Gyr} \lesssim t_{diss,4}$.  The irregular cluster formation history of the LMC, combined to completeness issues at low mass, prevent the exact pinning of the cluster mass function turnover carved by cluster evaporation and, therefore, a more accurate estimate of $t_{diss,4}$ for the LMC.         

The reader may comment that cluster age differences could account for the different SMC and LMC cluster locations in Fig.~\ref{fig:mallf1p}.  The LMC clusters shown in Fig.~\ref{fig:mallf1p} are indeed younger than their SMC counterparts (in Table~2 of \citealt{mil20}, LMC clusters are 1.6-2\,Gyr old, while SMC clusters are 1.6-10.5\,Gyr).  Age alone cannot explain the shift between the SMC and LMC clusters in the $(m_{prst}, F_{1P})$ space, however.  To show this, we have added to the sample of \citet{mil20} five LMC \gcs studied by \citet[][NGC~2210 and Hodge~11]{gil19} and \citet[][NGC~1466, NGC~1841 and NGC~2257]{gil20}, all of them older than 12\,Gyr \citep{bau13}.  Each of their main sequence presents a secondary population whose color indices $F336W-F606W$ and $F336W-F814W$ are redder than for the rest of the cluster stars \citep{gil19,gil20}.  This points to a nitrogen enhancement \citep{gil19}, because the $F336W$ filter covers a NH absorption band \citep[Fig.~1 in][]{pio15}.  These secondary stars should thus correspond to the polluted population.

\begin{figure} 
\begin{center} 
\includegraphics[width=0.49\textwidth, trim={0.cm 0.cm 0cm 0cm}]{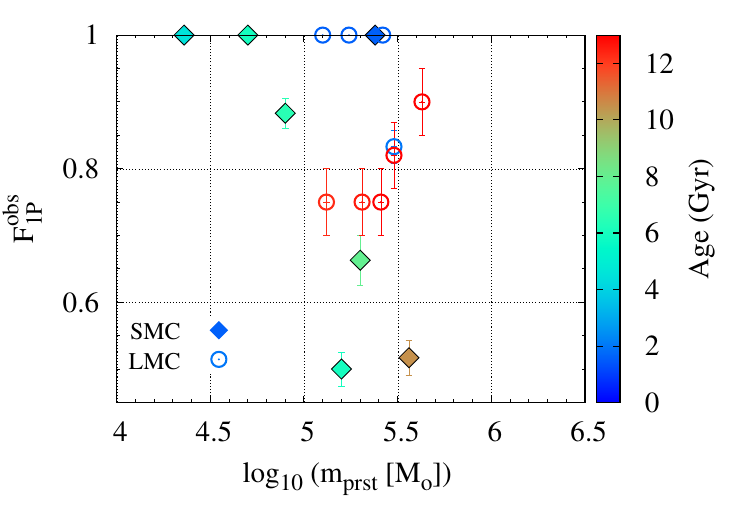} \\
\includegraphics[width=0.49\textwidth, trim={0.cm 0.cm 0cm 0cm}]{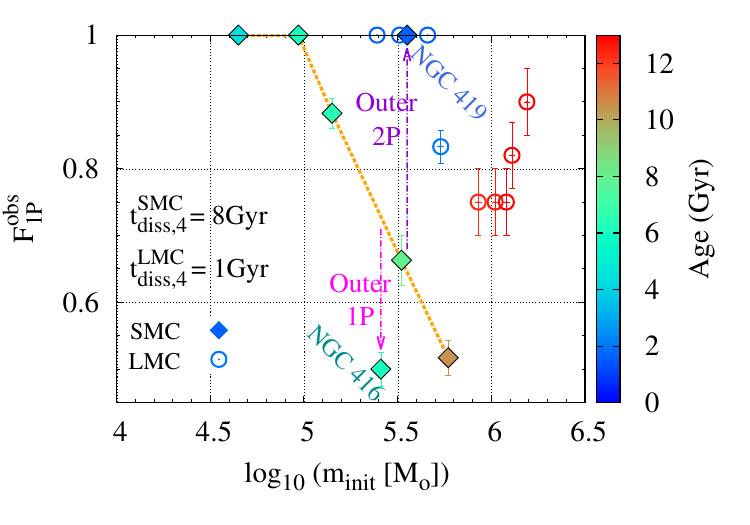}
\caption{The pristine-star fraction vs the present-day mass (top panel) and vs the initial mass (bottom panel) for the 5 ancient LMC clusters studied by \citet{gil19,gil20} (red open circles), and the LMC and SMC clusters studied by \citet{mil20} and already shown in Fig.~\ref{fig:mallf1p}.  SMC clusters are depicted as filled diamonds, LMC clusters as open circles.  Symbols are color-coded for cluster age (see the right-hand-side palette).  In the bottom panel, the LMC and SMC cluster dissolution time-scales used to estimate the cluster initial masses (see Eq.~\ref{eq:minit_lam05}) are quoted and a thick dashed orange line highlights the "spine" of the SMC initial sequence.  We speculate that the SMC clusters NGC~416 and NGC~419 lost preferentially their 1P and 2P stars, respectively, and the vertical dashed-dotted arrows illustrate their assumed pristine-star fraction evolutions following gas expulsion.  We therefore speculate that NGC~419 may have formed as a multiple-population cluster (see Sec.~\ref{sec:spmp}) 
}
 \label{fig:mag} 
 \end{center} 
\end{figure}

The top panel of Fig.~\ref{fig:mag} gathers in the $(m_{prst},F_{1P})$ space the samples of \citet{mil20}, \citet{gil19} and \citet{gil20}.  SMC clusters are depicted as filled diamonds, LMC clusters as open circles.  Symbols are color-coded for cluster age.  The present-day mass and age estimates for the newly-added LMC clusters are from \citet{bau13}\footnote{\url{https://cdsarc.cds.unistra.fr/ftp/J/MNRAS/430/676/table2.dat}}.  For Hodge~11 and NGC~2210, the secondary population amounts to 10\,\% and 18\,\%, respectively, of the entire main-sequence population \citep[after accounting for binary contamination; ][their conclusions]{gil19}\footnote{These fractions are not those quoted in their abstract.  We have opted for the fractions mentioned in their conclusions because (1)~they are in line with the information tied to their Figs~7-8 and (2)~they have been corrected for binary contamination.}.  We therefore adopt $F_{1P}=0.90$ for Hodge~11 and $F_{1P}=0.82$ for NGC~2210.  As for NGC~1466, NGC~1841 and NGC~2257, the secondary population represents 30-40\,\% of the whole main-sequence population, while the binary contamination is estimated to be $\lesssim 10$\,\% \citep{gil20}.  For these three clusters, we therefore use a pristine star fraction $F_{1P} \simeq 1 - (0.35-0.1) \simeq 0.75$.  As already found by \citet{mar18}, clusters younger than 1.7\,Gyr do not show sign of multiple populations.  However, single-population clusters can be older than 1.7\,Gyr \citep[e.g. Lindsay~38 and Lindsay~113;][]{mil20}

Despite their old age, the newly added LMC clusters remain to the right of their SMC siblings, even though they evolve in a stronger tidal field.  The situation is thus opposite to what has been observed for Galactic \gcs (\citealt{zen19,mil20}, Fig.~8 in Paper~I; see also the least-squares fits for inner- and outer-halo clusters in our Figs~\ref{fig:mtof1p}-\ref{fig:m2pf1p}).  We are thus left to conclude that the LMC and SMC clusters start from distinct initial sequences $F_{1P}(m_{init})$, that of the LMC being to the right of the SMC's.

This is illustrated in the bottom panel of Fig.~\ref{fig:mag}, where the cluster initial masses have been estimated based on the dissolution time-scales $t_{diss,4} \simeq 8$\,Gyr for the SMC \citep{lam05a} and $t_{diss,4} \simeq 1$\,Gyr for the LMC \citep{par08}.  These time-scales having been obtained under the formalism of \citet{lam05b}, we use their expression of the cluster-mass time evolution to recover the cluster initial masses.  Eq.~\ref{eq:lam05} below is their Eq.~6 rewritten with the tidal field strength quantified via the dissolution time-scale $t_{diss,4}$ \citep[see also Eqs~1 and 2 in][]{par08}
\begin{equation}
m_{cl}(t) = m_{init} \left[ \mu_{ev}(t)^\gamma 
                          - \frac{\gamma t}{t_{diss,4}} \left( \frac{10^4\,\Ms}{m_{init}} \right)^\gamma              
  \right]^{1/\gamma}  \;. 
\label{eq:lam05}
\end{equation}
Here, $t$ is the cluster age, $m_{cl}(t)$ is the cluster mass at age $t$, $\mu_{ev}(t)$ describes the time-dependent stellar evolution mass losses, and $\gamma=0.62$ \citep{lam05b}.  Eq.~\ref{eq:lam05} also provides a good match to the results of \citet{bm03} N-body simulations.
The clusters of our sample being too old to still experience any significant stellar evolutionary mass losses (age $\geq 1.6$\,Gyr), we adopt $\mu_{ev}(t) = F_{StEv}=0.7$ as previously.  In Table~2 of \citet{bau13}, some old LMC clusters are assigned an age of $\simeq 15$\,Gyr.  For such clusters, we adopt $t=13$\,Gyr. 

The initial mass $m_{init}$ of a cluster can now be expressed as
\begin{equation}
m_{init} = \frac{ \left( m_{prst}^\gamma 
                          + \gamma 10^{4\gamma}t / t_{diss,4} \right)^{1/\gamma}}{F_{StEv}}           
\label{eq:minit_lam05}
\end{equation}
Our results are presented in Table~\ref{tab:mag} and in the bottom panel of Fig.~\ref{fig:mag}.  The latter confirms that most LMC and SMC clusters draw two separate sequences in the $(m_{init},F_{1P})$ space, with the SMC sequence to the left of the LMC's\footnote{Note that a LMC cluster dissolution time-scale longer than its lower limit (i.e. $t_{diss,4} > 1$\,Gyr) would move the LMC points to masses intermediate between those of the top and bottom panels.}.  That the SMC clusters are located leftward of the LMC clusters in the $(m_{prst},F_{1P})$ space (top panel of Fig.~\ref{fig:mag}) must thus result from distinct formation properties and early evolutionary paths.
         
\begin{table}
\begin{center}
\caption{Cluster initial masses $m_{init}$ as estimated from Eq.~\ref{eq:minit_lam05} \citep[based on][]{lam05b}.  The cluster age and present-day masses are taken from Table~2 in \citet{mil20}, except for the old LMC clusters Hodge~11, NGC~2210, NGC~1466, NGC~1841, NGC~2257, whose age and present-day masses are taken from Table~2 of \citet{bau13}  } 
\begin{tabular}{l c l c } \tableline 
\multicolumn{4}{c}{SMC clusters: $t_{diss,4} = 8\,Gyr$}   \\ \tableline
ID             & $log( \frac{m_{init}}{\Ms})$ & ID & $log( \frac{m_{init}}{\Ms})$  \\ \tableline
Lindsay~1      & 5.52       &  NGC~121    & 5.77  \\   
Lindsay~38     & 4.97       &  NGC~339    & 5.15  \\     
Lindsay~113    & 4.65       &  NGC~416    & 5.41   \\
               &            &  NGC~419    & 5.55   \\ \tableline
\multicolumn{4}{c}{LMC clusters: $t_{diss,4} = 1\,Gyr$}   \\ \tableline
ID             & $log( \frac{m_{init}}{\Ms})$ & ID & $log( \frac{m_{init}}{\Ms})$  \\ \tableline
NGC~1783     &   5.66       &  Hodge~11  &  6.19  \\
NGC~1806     &   5.39       &  NGC~2210   &  6.11  \\
NGC~1846     &   5.51       &  NGC~1466   &  6.02  \\ 
NGC~1978     &   5.73       &  NGC~1841   &  5.93  \\ 
             &              &  NGC~2257   &  6.08  \\ \tableline
\end{tabular}
\label{tab:mag}
\end{center}
\end{table}

What "fragilizes" the SMC clusters compared to their LMC counterparts?  In the SMC, we expect clusters to form out of a lower-density environment than in the LMC.  As a result, SMC cluster progenitors may achieve lower star formation efficiencies \citep{par13,pol24}, which are themselves conducive to lower bound fractions by the end of violent relaxation (Fig.~15 in \citealt{par13}, Fig.~8 in \citealt{shu17}, Fig.~2 in \citealt{farias17}).  Clusters in the SMC may thus achieve initial masses lower than in the LMC for a given $F_{1P}$, in line with the bottom panel of Fig.~\ref{fig:mag}.

More detailed modeling and the observational mapping of more clusters are desirable.  This holds all the more true for the LMC since its cluster dissolution time scale is known as a lower limit only \citep{par08}, while, as stressed by \citet{gil20}, its old clusters still await their full characterisation with the "magic trio" of UV filters \citep[Fig.~1 in][]{pio15}.  It will then be interesting to see whether the spines of the LMC and SMC sequences $F_{1P}(m_{init})$ are parallel to each other.

\section{Did the single-population clusters Rup~106 and NGC~419 form as multiple-population clusters?}
\label{sec:spmp}

In the bottom panel of Fig.~\ref{fig:mag}, we have highlighted with a thick dashed orange line the "spine" of the SMC sequence in the $(m_{init},F_{1P})$ space.   This spine is already noticeable in the top panel (i.e. $(m_{prst},F_{1P})$ space) since, given the weakness of the SMC tidal field, stellar evolution mass losses dominate the shifts in cluster mass, which are thus about constant (i.e.~$\Delta log_{10}F_{StEv} = -0.15$).  Two SMC clusters strongly depart from the initial sequence, namely, NGC~416 and NGC~419.  Regarding NGC~416, a possible explanation is that its 2P population formed centrally concentrated, leading to the preferential tidal stripping of its 1P population once the cluster expands following gas expulsion.  The preferential removal of the 1P stars  is a key aspect of the AGB scenario, in which the cluster polluters are aymptotic giant branch stars of the first population \citep[see e.g.][]{der08,ves10}.  However, \citet{lei23} show that this is sometimes the pristine population that forms centrally-concentrated (their Fig.~15).  The reverse scenario should thus also be possible, namely, formation of a centrally-concentrated 1P population, resulting in the preferential removal of the cluster 2P stars and the rise of the pristine-star fraction.  With NGC~419, we take this latter scenario to the extreme, suggesting that the cluster expansion in the aftermath of gas expulsion was strong enough for almost all its 2P stars to be tidally removed.  In contrast to NGC~416 and NGC~419, the clusters defining the SMC "spine" in the $(m_{init},F_{1P})$ space would have evolved at about constant $F_{1P}$ (i.e. their formation yielded well-mixed pristine and polluted populations).     

Another intriguing single-population cluster is Rup~106, a Galactic outer-halo cluster and the most massive single-population cluster of Fig.~\ref{fig:mtof1p}.  Rup~106 shows up as an oddity in Fig.~4 of \citet{hua24}, where its metallicity and compactness (defined as the ratio between its initial mass and half-mass radius) qualify it as a multiple-population cluster.  Did Rup~106 form with an outer, subsequently tidally-stripped, polluted population?  Compared to the other seven single-population clusters of our sample, Rup106 has also the second shortest pericentric distance, $D_{per}=4.63$\,kpc \citep{bau19}\footnote{Among the eight single-population clusters of our sample, the pericentric distance of Rup~106 is second only to that of the not so compact cluster Pal~14}.  A short pericentric distance creates favourable conditions for the tidal removal of cluster outer regions.  

We therefore speculate that some multiple-population clusters eventually evolve into pristine-star clusters, provided the following conditions are  met: (1)~formation of the pristine population in the cluster inner regions, where it is shielded against the external tidal field, (2)~formation conditions leading to a strong cluster expansion (e.g.~because of a low star formation efficiency), thereby exposing the outward polluted population to the external tidal field, and (3)~before the two populations get mixed, the tidal stripping of the outer polluted population.  

Although more data are required before drawing solid conclusions, we note that Rup~106 and NGC~419 are the most massive single-population clusters of our Milky-Way and SMC samples, respectively.   Mapping such clusters over wider field-of-views than has been done so far - possibly targeting tidal tails - is desirable to investigate whether some 2P stars may have lingered in their outskirts.  This would prove that some pristine-star clusters started their life as multiple-populations clusters.

\section{Summary and conclusions}
\label{sec:conclu}

In Galactic globular clusters, the pristine-population fraction $F_{1P}$ is a strongly decreasing function of the polluted-population mass $m_{2P,prst}$.  $F_{1P}$ correlates less tightly with the cluster total mass $m_{prst}$, and only poorly with the pristine-population mass $m_{1P,prst}$.  According to \citet{mil20}, the correlation coefficients are $-0.77$, $-0.64$ and $-0.40$ in the $(m_{2P,prst},F_{1P})$, $(m_{prst},F_{1P})$ and $(m_{1P,prst},F_{1P})$ spaces, respectively (see their Fig.~7).  

We have deciphered what these distinct correlations can tell us about the formation of multiple stellar populations in Galactic globular clusters.  To this end, we have extended the mapping of the pristine-star fraction versus present-day cluster mass perfomed in \citet[][Paper~I]{par24}.  

Our main finding is: for a given $({\rm mass}, F_{1P})$ space (where "mass" is either $m_{2P}$, $m_{cluster}$ or $m_{1P}$), the correlation shown by present-day globular clusters reflects the slope of the corresponding initial track, $F_{1P}(m_{2P,init})$, $F_{1P}(m_{init})$ or $F_{1P}(m_{1P,init})$.  A steep initial track yields a loose correlation.  
For instance, that the present-day cluster data are weakly correlated in the $(m_{1P},F_{1P})$ space can be explained by an initial track that is vertical, i.e. the initial mass $m_{1P,init}$ in pristine stars is about constant (top panel in Fig.~\ref{fig:mallf1p}).  Here, "initial" refers to the onset of the cluster long-term secular evolution, equivalently the evolutionary stage when they have returned to virial equilibrium following gas expulsion.   

In greater detail, our results are as follow.  Sec.~\ref{sec:mtof1p} presents the updated cluster data set and summarizes Paper~I.  Its key hypotheses are: (a)~clusters contain a fixed mass $m_{th}$ in pristine stars by the end of their formation, and (b)~the pristine-star fraction $F_{1P}$ stays constant as clusters evolve up to their present age (i.e. pristine and polluted stars are lost equally likely).  The pristine-star fraction of a cluster at any stage of its evolution therefore obeys $F_{1P}=m_{th}/m_{ecl}$ (Eq.~\ref{eq:f1p-ecl}, magenta line in Fig.~\ref{fig:mtof1p}), with $m_{ecl}$ the cluster stellar mass at the end of its formation (i.e. straight before residual \sfing gas expulsion).  As for the initial track, it obeys Eq.~\ref{eq:f1p-init} (orange line in Fig.~\ref{fig:mtof1p}) for a fixed fraction $F_{bound}^{VR}$ of stars retained after gas expulsion.  The correlation coefficient for the 56 Type~I Galactic \gcs of our sample is $r_I = -0.56 \pm 0.10$.  

Sections~\ref{sec:m1pf1p} and \ref{sec:m2pf1p} convert the tracks of Paper~I/Sec.~\ref{sec:mtof1p} (i.e.~$F_{1P}$ in dependence of the cluster {\it total} mass) into tracks for the $(m_{1P},F_{1P})$ and $(m_{2P},F_{1P})$ spaces, respectively, with $m_{1P}=F_{1P} m_{cluster}$ and $m_{2P}=(1-F_{1P}) m_{cluster}$.  The respective initial tracks obey Eqs~\ref{eq:m1pf1p} and \ref{eq:m2pf1p} (orange lines in Figs~\ref{fig:m1pf1p}-\ref{fig:m2pf1p}).  For the 56 Type~I clusters, the correlation coefficient is $r_I = -0.26 \pm 0.13$ in the $(m_{1P},F_{1P})$ space and $r_I = -0.70 \pm 0.07$ in the $(m_{2P},F_{1P})$ space.  We therefore confirm the order of the correlations inferred by \citet{mil20}, namely, the pristine-star fraction is most tightly correlated with the polluted-population mass, and the correlation is the loosest with the pristine-population mass.    That our correlation coefficients are slightly weaker than \citet{mil20} 's stems from us excluding the Type~II clusters as their formation mechanism may differ from that of Type~I clusters.      

Section~\ref{sec:nocor} and its Fig.~\ref{fig:mallf1p} highlight with a yellow background the permitted region of each parameter space, thereby providing a direct vizualisation of how tight or how loose the corresponding correlation is.  The permitted region stretches from the initial track corrected for stellar-evolution mass losses (green track in the top panels of Figs~\ref{fig:mtof1p}-\ref{fig:m2pf1p} and in Fig.~\ref{fig:mallf1p}) to the cyan tracks, which mark the region where dynamical friction removes massive clusters.  A widely open permitted region (vertical initial track, top panel/$F_{1P}(m_{1P})$ in Fig.~\ref{fig:mallf1p}) corresponds to a poor correlation, while a narrower permitted region (shallow initial track, bottom panel/$F_{1P}(m_{2P})$ in Fig.~\ref{fig:mallf1p}) indicates a stronger correlation.  Therefore, our model accounts successfully for the decreasing strength of the observed  (anti)correlations, from the $(m_{2P},F_{1P})$ space, down to the  $(m_{cluster},F_{1P})$ and $(m_{1P},F_{1P})$ spaces.  

How does the model respond if the initial mass in pristine stars is not fixed, but rather a decreasing function of $F_{1P}$  (Eq.~\ref{eq:m1pf1p_sh} with $\psi < 1$)?  Figure~\ref{fig:mallf1p_sh} demonstrates that the pristine-star fraction must then correlate more strongly with all three masses $m_{1P,prst}$, $m_{prst}$ and $m_{2P,prst}$ (i.e. the permitted / yellow-highlighted regions in Fig.~\ref{fig:mallf1p_sh} shrink compared to Fig.~\ref{fig:mallf1p}).  At first glance, this may be seen as contradicting the poor correlation coefficient shown by Type~I clusters in the $(m_{1P}, F_{1P})$ space ($r_I = 0.26 \pm 0.13$; Sec.~\ref{sec:m1pf1p}).  Yet, for now, we will not exclude a scenario built on shallower initial tracks, like  those in Fig.~\ref{fig:mallf1p_sh}.  The reason is that the permitted regions in Figs~\ref{fig:mallf1p} and \ref{fig:mallf1p_sh} stem from an idealized scenario in which the relations between $F_{1P}$ and $(m_{2P}, m_{cluster}, m_{1P})$ are one-to-one, that is, devoid of scatter (orange tracks in all panels).  This is unlikely to be the case at the end of the gas-expulsion-driven cluster expansion (i.e. end of violent relaxation; transition from the magenta to the orange tracks).  At that stage, we do expect the actual initial distribution of \gcs to show some scatter.  Specifically, we expect (1)~a horizontal scatter due to the many parameters influencing $F_{bound}^{VR}$, and (2)~a vertical scatter driven by variations of the pristine-star fraction when one population is less centrally concentrated, hence more tidally-stripped, than the other.  Such a scatter would blur the neat edges of the permitted regions of Figs~\ref{fig:mallf1p} and \ref{fig:mallf1p_sh}, thereby weakening each correlation.  On top of their age and mass, the metallicity of clusters contribute to the onset of their multiple populations as well \citep[see Fig.~4 in][]{hua24}.  Cluster-to-cluster metallicity variations should thus also contribute to the initial scatter in any $({\rm mass},F_{1P})$ space.  

Section~\ref{sec:mag} stresses that the Magellanic Clouds clusters behave unexpectedly, the SMC clusters being located to the left of their LMC counterparts in the $(m_{prst},F_{1P})$ space (top panel of  Fig.~\ref{fig:mag}).  Given the weaker tidal field of the SMC, we would have expected the opposite, similar to what has been observed for the outer- and inner-halo \gcs of our Galaxy (see least-squares fits in bottom panels of Figs~\ref{fig:mtof1p}-\ref{fig:m2pf1p}; see also \citealt{zen19,mil20}).  We have estimated the initial masses of the LMC and SMC clusters in our sample (Table~\ref{tab:mag}).  This confirms that the LMC and SMC clusters draw two distinct sequences in the $(m_{init},F_{1P})$ space, with, for a given $F_{1P}$, the SMC clusters being on average less massive than their LMC counterparts, already initially (bottom panel of Fig.~\ref{fig:mag}).  We suggest this effect to stem from forming clusters achieving lower star formation efficiencies in the low-density SMC than in the LMC, yielding in turn larger cluster mass losses after gas expulsion in the SMC than in the LMC.

Finally, Sec.~\ref{sec:spmp} suggests that some single-population clusters (e.g. NGC~419 in the SMC and Rup~106 in the Galactic halo) formed as multiple-population clusters \citep[see also Fig.~4 in][]{hua24}.  The rise of their pristine-star fraction would follow from the severe tidal stripping of their outer 2P population after gas expulsion.

Based on Secs~\ref{sec:mag} and \ref{sec:spmp}, we expect the initial distribution of clusters in the $(m_{init},F_{1P})$ space to show some scatter.  Specifically, we expect (1)~a horizontal scatter, driven by the variations of $F_{bound}^{VR}$, the post-violent-relaxation bound fraction, and (2)~a vertical scatter, driven by variations of $F_{1P}$, the pristine-star fraction.  We will soon have the opportunity to revisit both aspects.        

\section*{acknowledgments} 

\small

GP acknowledges funding by the Deutsche Forschungsgemeinschaft (DFG, German Research Foundation) -- Project-ID 515414180.  The initial idea of Secs~\ref{sec:mtof1p}-\ref{sec:m2pf1p} came into being during SFB881, funded by the Deutsche Forschungsgemeinschaft -- Project-ID 138713538, SFB 881 ("The Milky Way System", subproject B05, PI: Dr.~A.~Pasquali).  GP is grateful to the anonymous referee for helpful suggestions.  This research has made use of NASA 's Astrophysics Data System.


\bibliography{ms}{}
\bibliographystyle{aasjournal}



\end{document}